\documentclass[a4paper,twocolumn]{revtex4}  
\usepackage{latexsym}
\usepackage[colorlinks=true,linkcolor=blue,anchorcolor=blue,citecolor=blue]{hyperref}
\usepackage[hmargin={1cm,1cm}, vmargin={2cm,2cm}]{geometry}
\usepackage{float}
\usepackage{graphicx,amsfonts,amsmath,amssymb,amsthm,amscd}
\usepackage{epsfig}
\usepackage{bm}
\usepackage{xr}
\usepackage{indentfirst}
\usepackage{appendix}
\usepackage{changepage}
\usepackage{color}

\date{\today}

\begin{document}



\title{Cooperation of myosin II in muscle contraction through nonlinear elasticity}

\author{Beibei Shen and Yunxin Zhang} \email[Email: ]{xyz@fudan.edu.cn}
\affiliation{Shanghai Key Laboratory for Contemporary Applied Mathematics, School of Mathematical Sciences, Fudan University, Shanghai 200433, China.}

\begin{abstract}
	Myosin II plays a pivotal role in muscle contraction by generating force through the cooperative action of multiple motors on actin filaments. 
	In this study, we integrate the nonlinear elasticity of the neck linker in individual myosin II and comprehensively investigate the evolution of cooperativity and dynamics at {\it microstate} and {\it mesostate} levels using a combined model of single and multiple motors.
	We find that a substantial proportion of actin-bound motors reside in the {\it mid-} and {\it post-power stroke} states, and our nonlinear model reveals their increased capacity for load sharing.
    Additionally, we systematically explore the impact of mechanical load and ATP concentration on myosin II motors. Notably, we observe that the average net distance of actin undergoes a transition from a weak load-sensitive regime at low ATP concentrations to a load-sensitive regime at higher ATP concentrations. Furthermore, increasing the load or raising the ATP concentration to saturation can enhance the efficiency and output power of myosin filament. Moreover, the efficiency of the myosin filament increases with the power stroke strength, reaching a maximum at a specific range, and subsequently declining beyond that threshold.	
	Finally, we explore the mean run time/length and mean existence probability of myosin filament, shedding light on its overall behavior.
\end{abstract}
	
	
	
\maketitle

\section{Introduction}
In cells, molecular motors such as myosin, kinesin, and dynein play essential roles in powering cellular activities, and their coordinated efforts are critical for the proper functioning of various biological systems, including muscle contraction, vesicle transport, and spindle formation \cite{howard2001mechanics, mcgrail1997microtubule, metzger2012map, wilson2015nesprins}. 

The myosin II family, encompassing skeletal, cardiac, smooth, and non-muscle myosins, can interact with actin filaments, and ultimately result in actin filaments sliding and macroscopic force production \cite{erdmann2016sensitivity,dayraud2012independent}. The mechanical properties of myosin II have been studied at the single-molecule level extensively, with a predominant focus on the optimization as an individual motor \cite{finer1994single, molloy1995movement,kaya2010nonlinear,ruegg2002molecular,houdusse2016myosin,elangovan2012integrated}. However, the generation of muscle contraction requires the coordinated efforts of multiple myosin II motors to pull actin filaments \cite{wagoner2021evolution,kaya2017coordinated,smith2003cooperative}. In this context, myosin II interacts with actin filaments not as independent force generators, but rather as cooperative force generators \cite{huxley1957muscle, huxley1971proposed,caremani2015force}.

Recently, the collective behaviors of multiple myosin II in vitro have been studied deeply to understand how individual motor functions are integrated into collective motor systems and contribute to the overall performance of cooperative motor systems \cite{piazzesi2007skeletal, linari2015force,stam2015isoforms,linari2009effect,irving2000conformation,lan2005dynamics}. But so far the cooperation mechanism among multiple myosin II in muscle contraction remains elusive. Meanwhile, the system of myosin II might have been evolved with distinct mechano-kinetic properties, which maximize output power or energy efficiency. These properties remain incompletely characterized and warrant further investigation.

To fully understand the collective behaviors of motor proteins, it is crucial to consider both the single-molecule properties and the intricate mechanochemical and/or chemical interactions among motor proteins \cite{badoual2002bidirectional,guerin2010coordination}. In this study, we will try to discover the cooperation mechanism among myosin II motors by taking both the {\it microstate} and {\it mesostate} of the myosin II system into account. In view of this, we will employ a model that allows for the detection and analysis of the dynamic behavior of individual myosin II, as well as their ensemble effects. Different with previous study, our model incorporates the nonlinear elasticity of the neck linker in myosin II, which closely approximates the mechanical behavior of myosin during muscle contraction \cite{kaya2010nonlinear, wagoner2021evolution}. Furthermore, we investigate the intricate effects of mechanical load and ATP concentration on the functionality of myosin II system.

\section{Theoretical description of muscle contraction}
\subsection{The mechanochemical cycle of single myosin II}
\begin{figure}[htbp]
 \includegraphics[scale=0.12]{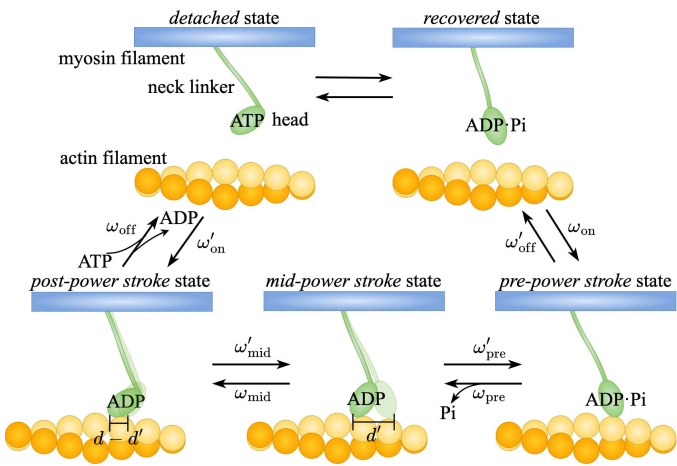}\\
	\caption{\label{single_cycle}
		The mechanochemical cycle of single active myosin II. The myosin head incorporates a lever arm and is connected to the backbone of myosin filament via a neck linker. For simplicity, we draw the lever arm and myosin head together as an oval shape. Actin filaments are represented by a double helix structure. The cycle of interaction between the myosin 
		and actin starts from the {\it recovered} state. During ATP hydrolysis, the myosin binds to actin, pulls actin against the external load, and then detaches. The lever arm and neck linker play crucial roles in this process, allowing the myosin to undergo conformational changes and perform the power stroke, which drives the movement of actin filaments. $\omega$ is the transition rate of motor between two states.}
\end{figure}
According to previous models of  actomyosin \cite{kaya2017coordinated, wagoner2021evolution,erdmann2016sensitivity,marcucci2016including}, we describe the work cycle of myosin II by a mechanochemical model with five states, see {\bf Fig.~\ref{single_cycle}}. 
The cycle begins with the unbound, \textit{recovered} state, where the motor head is loaded with ADP and phosphate (Pi), and the lever arm is in a primed conformation.
Notably, the head of myosin II possesses a lever arm that amplifies even the slightest conformational changes resulting from the binding or dissociation of various ligands at the nucleotide binding site \cite{duke1999molecular}.
Subsequently, the head of myosin II binds to actin in a \textit{pre-power stroke} state, which is a weakly bound state and does not generate force \cite{houdusse2016myosin}. Once the head attaches to actin, the rapid release of Pi triggers a substantial rotation, known as the \lq\lq power stroke", of the lever arm \cite{duke1999molecular,kaya2010nonlinear}.
As the lever arm swings to its stretched conformation, the motor executes power stroke, thereby increasing the strain on  neck linker and enhancing the force generated by the motor. Throughout execution of power stroke, the lever arm continues to rotate, pulling actin filament forward against the external load $F$.

During power stroke, the lever arm swings forward by a distance of $d = 8$ nm \cite{duke1999molecular,vilfan2003instabilities}. The power stroke comprises two distinct stages, marked by transitions towards \textit{mid-power stroke} and \textit{post-power stroke} states. 
During the first stage, the motor strain elongates by $d^{\prime}$ nm, while in the second stage, the strain elongates by $d-d^{\prime}$ nm. Generally,  the elongation of the strain in the first stage is slightly greater, ranging from 4 to 8 nm \cite{kaya2017coordinated}.

After power stroke, ADP dissociates from myosin II with a slow rate \cite{greenberg2016perspective,caremani2015force}, which is immediately followed by the binding of a ATP molecule. The binding of ATP destabilizes the actomyosin interaction, causing the myosin head to dissociate from actin filament and complete the cycle \cite{duke1999molecular}.
Finally, along with ATP hydrolysis, myosin II returns to the beginning of the cycle, that is, to the \textit{recovered} state again.

As illustrated in {\bf Fig.~\ref{single_cycle}}, the transition rates between these five states are influenced by both ATP concentration [ATP] and external load $F$. Specifically, the transition rates $\omega_{\rm on}$, $\omega_{\rm pre}^{\prime}$ and $\omega_{\rm mid}^{\prime}$ are solely dependent on the concentration of ATP, while the rate $\omega_{\rm off}^{\prime}$ is exclusively influenced by load $F$. The remaining transition rates $\omega_{\rm pre}$, $\omega_{\rm mid}$, $\omega_{\rm off}$ and $\omega_{\rm on}^{\prime}$ are affected by both ATP concentration and load $F$. The cycle of a single myosin II motor is actually mirrored by the cycle of ATP hydrolysis.	The specific expressions for each  transition rate can be found in Section E of the {\bf Supporting Material}.

We should also focus on the change in basic free energy across the five transitions. 
From the {\it recovered} state to the {\it pre-power stroke} state, the basic free energy of a myosin II motor undergoes a change of $g_{\rm on}$. 
When moving clockwise around the cycle from the {\it pre-power stroke} state, the basic free energy undergoes changes across the other transitions, which are labeled as $g_{\rm ps}/2$, $g_{\rm ps}/2$, $g_{\rm off}$, $g_{\rm recovery}$.       
Due to the changes in conformational free energy sum to zero across a full cycle, the sum of these components is equal to the free energy gained from ATP hydrolysis:
$g_{\rm on}+g_{\mathrm{ps}}+g_{\mathrm{off}}+g_{\text {recovery }}=-\Delta \mu_{\mathrm{ATP}},$
where $\Delta \mu_{\mathrm{ATP}}\geq0$, as it represents the chemical potential energy of the system.
The value of $\Delta \mu_{\mathrm{ATP}}$ is dependent on the concentration of ATP. 
 The formula is
\begin{equation}
	\Delta \mu_{\rm ATP}=25+\ln{\frac{[\rm ATP]}{2000}}.
	\label{ATPG}
\end{equation}
As ATP concentration increases, $\Delta\mu_{\rm ATP}$ also increases. At a standard concentration of $[\rm ATP]=2000\mu M$, the free energy change of ATP hydrolysis is 25 $k_{\rm B}T$, where $k_{\rm B}T$ is used as the unit of energy, $k_{B}$ is Boltzmann's constant and $T$ is the absolute room temperature.
\subsection{The cooperation mechanism among multiple myosin II motors}
The fundamental building block of muscle tissue is half-sarcomere, which is composed of overlapping sets of myosin and actin filaments \cite{piazzesi2007skeletal}. In muscle contraction, multiple myosin motors bind and output mechanical force on actin filaments cooperatively to produce relative sliding between myosin and actin filaments \cite{piazzesi2007skeletal,kaya2017coordinated,linari2015force}.
\begin{figure}[htbp]
	\includegraphics[scale=0.106]{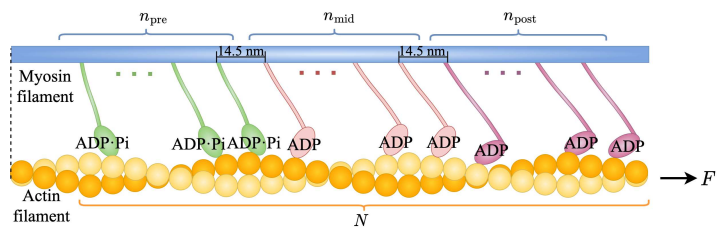}\\
	\caption{Schematic of $N$ myosin II motors interacting with actin filaments. The external load $F$ pulling to the right is balanced by forces in neck linkers of myosin II motors. The spatial interval between each two motors on the backbone of myosin filament is fixed at 14.5 nm \cite{piazzesi2007skeletal}. In muscle contraction, myosin II motors must act cooperatively. To increase the possibility of coordinated power stroke, motors need to be in \textit{pre-power stroke} state or \textit{mid-power stroke} state transiently.
	In our model, we assume that among the $N$ bound myosin II motors, there are $n_{\rm pre}$ myosin II motors in \textit{pre-power stroke} state, $n_{\rm mid}$ motors in \textit{mid-power stroke} state, and the other $n_{\rm post}$ ones in \textit{post-power stroke} state. For more details of the coordinated {\it microstate} vector cycle, see {\bf Fig.~S2} in the {\bf Supplemental Material}.}
	\label{Fig:2}
\end{figure}

Multiple myosin motors are arranged into well-organized superstructures, see {\bf Fig.~\ref{Fig:2}}. 
In the structures, a vast assembly of myosin functions collectively as a single functional unit, operating in a coordinated manner to facilitate efficient relative sliding of actin and myosin filament \cite{piazzesi2007skeletal}. The backbone of myosin filament displays a pattern of myosin distribution, with clusters of myosin emerging at fixed 14.5 nm intervals \cite{piazzesi2007skeletal}. 
A myosin half-filament is found to have a total of $N_{0}=294$ myosin motors, each of which can be either active or inactive. The number of active myosin motors $N_{T}$ depends on load $F$, see Eq.~(S3) \cite{linari2015force}. 

The mechanochemical cycle of each active myosin motor consists of five possible states. Myosin motor in \textit{pre}-, \textit{mid}-, or \textit{post-power stroke} state binds to actin filament, while the one in \textit{detached} or \textit{recovered} state is separated from actin filament, as shown in {\bf Fig.~\ref{single_cycle}}. We consider both the {\it microstate} and the {\it mesostate} of this myosin system. The {\it microstate} is described by a vector ${\bf n}=\left(n_{\rm pre}, n_{\rm mid}, n_{\rm post}, n_{\rm det}, n_{\rm rec}\right)$, where the subscript indicates the mechanochemical state and $n$ is the number of myosin motors in that state. The {\it mesostate} is described by the number $N$ of myosin motors bound to actin filament ($N=n_{\text{pre}}+n_{\text{mid}}+n_{\text{post}}$). So the number of active myosin motors detached from actin filament is $N_{T}-N=n_{\rm det}+n_{\rm rec}$.

Molecular properties of myosin II are specifically tuned to perform cooperative force generation for efficient muscle contractions \cite{wagoner2021evolution,linari2015force,kaya2017coordinated}. The bound myosin motors in {\it power stroke} states transit in a specific order. To describe the mechanism of multiple myosin motors which act cooperatively, we define the following sequence of vectors in {\it microstate} cycle,
\begin{equation}
	\begin{aligned}
		&\mathbf{n}=\left(n_{\text {pre}}, n_{\text {mid}}, n_{\text {post}}, n_{\text {det}}, n_{\text {rec}}\right), \\
		&\mathbf{n}^{\prime} =\left(n_{\text {pre}}, n_{\text {mid}}, n_{\text {post}}-1, n_{\text {det}}+1, n_{\text {rec}}\right), \\
		&\mathbf{n}^{\prime \prime} =\left(n_{\text {pre}}, n_{\text {mid}}, n_{\text {post}}-1, n_{\text {det}}, n_{\text {rec}}+1\right), \\
		&\mathbf{n}^{\prime \prime \prime} =\left(n_{\text {pre}}+1, n_{\text {mid}}, n_{\text {post}}-1, n_{\text {det}}, n_{\text {rec}}\right), \\
		&\mathbf{n}^{\prime \prime \prime \prime} =\left(n_{\text {pre}}, n_{\text {mid}}+1, n_{\text {post}}-1, n_{\text {det}}, n_{\text {rec}}\right),
	\end{aligned}
	\label{n5}
\end{equation}
with the {\it microstate} vector cycle given as (see {\bf Fig.~S2})
$$\mathbf{n}^{\prime \prime} \rightarrow \mathbf{n}^{\prime \prime \prime} \rightarrow \mathbf{n}^{\prime \prime \prime \prime} \rightarrow \mathbf{n} \rightarrow \mathbf{n}^{\prime} \rightarrow \mathbf{n}^{\prime \prime}.$$ 
During each cycle, one ATP molecule is hydrolyzed, and the actin filament is pulled forward with a certain distance. In the following, we will describe the {\it microstate} cycle in detail.

We assume the {\it microstate} cycle starts from $ \mathbf{n}^{\prime \prime}$, and the actin filament is bound with $N-1$ myosin motors. We give each bound myosin an index from left to right, indicating their order of duration bound with actin filament, where $i = 1$ represents the one most recently bound, and therefore in the {\it pre-power stroke} state, while $i=N-1$ represents the first bound one, and most probably in the {\it post-power stroke} state, see {\bf Fig.~S2}.

Subsequently, an additional motor binds to actin in the {\it pre-power stroke} state, thereby increasing the number of bound motors to $N$. This transition in the cycle leads to {\it microstate} $\mathbf{n}^{\prime \prime\prime}$.

To increase the likelihood of coordinated power strokes, multiple myosin motors must briefly \lq\lq stay" in either the {\it pre-power stroke} or {\it mid-power stroke} state \cite{kaya2017coordinated}. Compelling evidence from recent experimental results suggests that the stepwise displacements of actin are primarily generated through the coordination of power strokes among two myosin motors  \cite{kaya2017coordinated}. According to myosin motors transition in a first-in, first-out manner \cite{wagoner2021evolution}, the transition from ${\bf {n}'''}$ to ${\bf {n}''''}$ corresponds to motor $i = n_{\rm pre}$ completing the first half of the power stroke and transitioning to the {\it mid-post power stroke} state. 

During the first stage of the power stroke, the strain of motor $i = n_{\rm pre}$ elongates by $d^{\prime}$ nm, simultaneously exerting an increasing force. 
This disrupts the initial force balance relationship, resulting in the actin filament being pulled to the forward by a distance $\Delta x_{1}$.

After that, the transition from ${\bf {n}''''}$ to ${\bf n}$ corresponds to motor $i = (n_{\rm pre}+n_{\rm mid})$ completing the second stage of the power stroke and entering the {\it post-power stroke} state. During this phase, the rotation of the lever arm causes the strain of the $n_{\rm pre}+n_{\rm mid}$-th motor to elongate by $d-d^{\prime}$ nm. Similarly, motors pull the actin filament to the forward by a distance $\Delta x_2$. 

Then, the release transition from ${\bf n}$ to ${\bf{n}'}$ corresponds to the unbinding of the $i = N$ motor. When a {\it post-power stroke} state motor detaches from the actin filament, the force exerted by the bound motors suddenly decreases, breaking the previous force equilibrium. As a result, the actin filament slips backward by a distance of $\Delta x_{\rm slip}$, causing the strain of the other motors to be stretched until this remaining $N-1$ motors are equilibrated with the load $F$ again. This process is similar to when someone quits a tug-of-war, the rope immediately slides in the direction of the opposing team. See {\bf Fig.~S2} for illustration. 

Finally, a motor in the {\it detached} state transitions to the {\it recovery} state, which corresponds to the cycle transitioning from ${\bf{n}'}$ to ${\bf{n}''}$, ultimately bringing the cycle back to ${\bf{n}''}$. 

After such a cycle, the size of the net distance moved by the actin filament is $\Delta x_{\rm net}=\Delta x_{1}+\Delta x_{2}-\Delta x_{\rm slip}$. Here we only provide a simplified expression for $\Delta x_{\rm net}$ for the sake of clarity, as shown in {\bf Fig.~S2}. 
The specific formula with detailed elements can be found in the Method Eq.~\eqref{Xnet1} and Section G of the {\bf Supplemental Material}.

	\subsection{Nonlinear elasticity of myosin II }
In contrast to other models \cite{wagoner2021evolution,stam2015isoforms,duke1999molecular,erdmann2013stochastic, erdmann2012stochastic}, we propose that the mechanism of generating force during positively strained and negatively strained neck linker is different, and that the neck linker is not simply approximated as a linear spring model. Actually, experiments measuring the elastic properties of  myosin motors for a wide range of positive and negative strains demonstrate that myosin motors exhibit nonlinear elasticity \cite{kaya2010nonlinear}. Taking account of the nonlinear nature of myosin motor elasticity is essential to relate myosin's internal structural changes to physiological force generation and filament sliding. 

If myosin motors are under positive strain, they are considered to be \lq\lq stretched". In such cases, the neck linker can be approximated as a worm-like chain (WLC) model \cite{fang2021biomechanical,fedosov2010systematic,li2005spectrin}. On the other hand, if motors are under negative strain, they are referred to as \lq\lq drag" motors, exhibiting significantly lower stiffness compared to positively strained myosin motors. Low stiffness minimizes drag of negatively strained motors during muscle contraction at loaded conditions \cite{kaya2017coordinated}.
The nonlinear elasticity implies that active myosin motors with high stiffness is primarily responsible for the forward step generation, whereas the drag myosin motors with low stiffness does not contribute to forward \cite{kaya2010nonlinear}.

The force-extension relationship of a bound motor $i$ depends on the strain, $y_{i}$, as follows:
\begin{equation}
	f(y_{i})=
	\left\{\begin{aligned}
		&\frac{k_{B} T}{L_{p}}\left[ \frac{1}{4(1-(y_{i}/L_{C}))^{2}}-\frac{1}{4}+\frac{y_{i}}{L_{C}}\right],&y_{i} > 0 ,\\
		&\frac{k_{\mathrm{m}}}{\gamma}[\exp (\gamma \cdot y_{i})-1],& y_{i}\le 0.\\
	\end{aligned}\right.
	\label{fWLC}
\end{equation}
where $y_{i}$ represents the strain of motor $i$, and $f(y_{i})$ is a continuous function with continuous derivatives. The constant $k_{\mathrm{m}}$ is defined as $3k_{B}T/(2L_{p}L_{C})$. The dimensionless constant $\gamma$ is a parameter that characterizes the nonlinearity of the motor's response. The quantities $L_{C}$ and $L_{p}$ represent the contour length and persistence length of the motor chain, respectively, and they are both positive. The ratio $y_{i}/L_{C}$ corresponds to the extension of the motor, and it takes values in the range $(0,1)$.

{\bf Fig.~S1(a)} shows the nonlinear elastic force of the bound myosin motor $i$ as a function of the extending ratio $y_{i}/L_{c}$. When $y_{i}<0$, the slope of the curve is $k_{\rm m}\exp(\gamma\cdot y_{i})>0$, which decreases and becomes flatter as $\gamma$ increases. As $y_{i}$ becomes more negative, the resulting force $f(y_{i})$ rapidly approaches a negative constant $-k_{\rm m}/\gamma$, and the magnitude of this constant $|-k_{\rm m}/\gamma|$ decreases with increasing $\gamma$, indicating that the force $f(y_{i})$ decreases with increasing $\gamma$. 
When $y_{i}$ is positive and approaches zero, the behavior of the motor's neck linker is similar to that of a linear spring. However, as $y_{i}$ increases, the nonlinearity of the neck linker becomes more prominent. In contrast to the linear model, our nonlinear model reveals that motors in the {\it mid-} and {\it post-power stroke} states effectively distribute a larger load.

When $N$ motors attach to actin filament, the force exerted by these motors should be equal the external load $F$,
\begin{equation}
	\sum_{i=1}^{N} f(y_{i})=F.
	\label{equal_F}
\end{equation}
Eq.~\eqref{equal_F} illustrates the relationship between the external load $F$ and the forces generated by all motors. This force sensitivity enables any myosin motor to coordinate its actions with other bound myosin motors. As the external load $F$ increases, a greater number of motors are needed to counter balance it. This phenomenon is demonstrated by the experimental results shown in {\bf Fig.~3(b)}.

We calculate the change of motors' strain through mathematical derivation based on the first-in, first-out transitions of myosin motors. Meanwhile, we observed a self-consistent strain distribution at the end of {\it microstate} vector cycle, which was identical to that at the beginning of {\it microstate} cycle except for an increase in the indices of all bound motors by one and a movement of actin filament by a certain distance. The detailed derivation of which is presented in Section C of {\bf Supplemental Material}. 

Finally, we find that each microstate ${\bf n}''$ has a unique set of strain values, which enables us to capture the distribution of strain across the bound motors. Specifically, we can describe the strain of the bound motor $i$ as follows:
\begin{equation}
		y_{i}(\mathbf{n}'', F)=i \Delta y(\mathbf{n}'',  F)+d_{i}(\mathbf{n}''), 
	\label{yi}
\end{equation} 
where $i= 1, \cdots, N-1$, and $d_{i}(\mathbf{n}'') =0,d^{\prime}$ and $d$ for a motor in the {\it pre-power stroke}, {\it mid-power stroke}, and {\it post-power stroke} states, respectively. We assume that the strain on a pair of consecutive motors differs by a constant $\Delta y({\bf n}'', F)$.

\section{Results}
\subsection{Biophysics of myosin motion along actin filament}
We parameterize our model based on data from Piazzesi $et.\ al.$ \cite{piazzesi2007skeletal}, which provides information on the velocity $V$, the average number of bound motors $\left\langle N\right\rangle $, and the sliding distance $ L $ as functions of the external load $F$.
\begin{figure}[htbp]
	\includegraphics[scale=0.43]{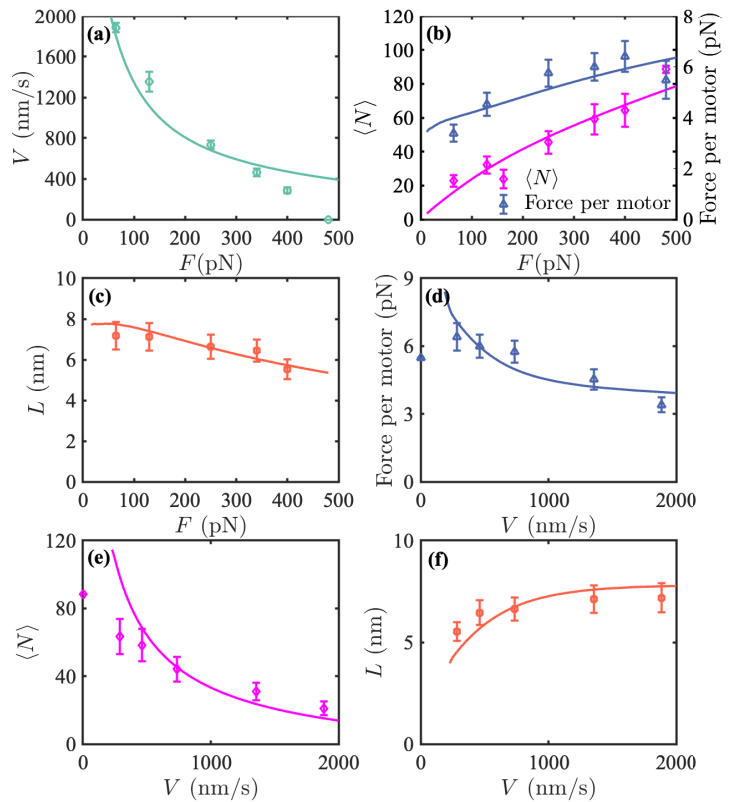}\\
	\caption{\label{fit_F} Theoretical predictions (solid lines) and experimental data (markers) of various biophysical properties of myosin II from muscles of frogs.
The data in (a-f) are from the study by Piazzesi $et.\ al.$ \cite{piazzesi2007skeletal}. The data in (a-c) are the original data used for model fitting, while the data in the (d-f) are obtained from the relationship between $\left\langle N\right\rangle $, $F/\left\langle N\right\rangle $, $V$, $L$ and $F$. $V$ is the velocity of the actin filament. In (b), the left axis is for average number of myosin II motors bound to actin $\left\langle N\right\rangle $, while the right axis is for force per attached motor $F/\left\langle N\right\rangle $. $L$ is the sliding distance of actin filament. Theoretical results are obtained from formulations given in Eqs. \eqref{N}, \eqref{V} and \eqref{L}, with model parameters listed {\bf Tab.~\ref{tab:1}}. The depicted data in this figure corresponds to a physiologically ATP concentration of 2000 $\mu$M.}
\end{figure}

Using parameter values listed in {\bf Tab.~\ref{tab:1}}, our model successfully reproduces the relevant experimental results shown in {\bf Fig.~\ref{fit_F}}. Detailed methods of theoretical prediction are presented in Method section and the Section G of {\bf Supplemental Material}. In the subsequent sections, we will consistently refer to the parameter values from {\bf Tab.~\ref{tab:1}} to investigate the mechanochemical properties of myosin II. In {\bf Fig.~\ref{fit_F}}, the myosin motors are assumed be in biological environment with ATP concentration of 2000 $\mu$m. In {\bf Fig.~S8}, we provide predicted curves of myosin II properties versus load $F$ at different ATP concentrations. Next, we will analyze the effect of load $F$ on myosin II motors by combining the information presented in both {\bf Fig.~\ref{fit_F}} and {\bf Fig.~S8}.

{\bf Fig.~\ref{fit_F}(a)} demonstrates a monotonic decrease of velocity with increasing load $F$, consistent with the downward convex velocity-force relationship at constant [ATP] described by the Hill relation \cite{hill1938heat}. On the other hand, the decrease in ATP concentration from the mechanosensitive regime, occurring at near-vanishing load, leads to a rapid reduction in velocity, as shown in {\bf Fig.~S8(a)}. However, as the load increases, the rate at which velocity decreases gradually slows down. This phenomenon has been thoroughly investigated in skeletal muscle and previously examined in motility assays \cite{walcott2012mechanical}.

With the increases of  load $F$, the number of bound motors $\left\langle N\right\rangle $ exhibits an increasing sensitivity to variations in ATP concentration, see {\bf Fig.~S8(b)}.
The load $F$ is shared by all myosin motors bound to actin filament. {\bf Fig.~\ref{fit_F}(b)} shows that both the average number $\left\langle N\right\rangle $ of myosin motors bound to actin filament and the average force shared by each myosin, defined as $F/\left\langle N\right\rangle $, increase with load $F$. We may need to point out that, the load shared by each myosin motor are actually different, and can be obtained by Eq.~(\ref{equal_F}). 

Next, we introduce the sliding distance $ L $ of actin filament. 
During each {\it microstate} cycle, the actin filament is displaced by a net distance $\Delta x_{\text {net}}$, which can be referred to as the net step size of the actin, and we denote $\left\langle \Delta x_{\text {net}}\right\rangle $ as average net distance. So the output work of the total myosin ensemble is $F\Delta x_{\rm net}$. As in \cite{piazzesi2007skeletal}, the sliding distance $L$ of actin filament is calculated as work divided by the force per motor $F/N$. 
This motivates the relation 
\begin{equation}
 L =\left\langle N\Delta x_{\text {net}}\right\rangle.
	\label{L}
\end{equation}
The distance that the actin filament slides from one motor attachment to detachment can be interpreted as the sliding distance. In other words, $L$ represents the distance that the actin filament slides while a motor remains attached to it \cite{piazzesi2007skeletal, wagoner2021evolution}. 

The sliding distance $L$ gradually decreases with increasing load $F$, as shown in {\bf Fig.~\ref{fit_F}(c)}. However, the  load $F$ has a limited effect on the sliding distance $L$, which remains around 6 nm at large loads and 8 nm at small loads. 
On the other hand, when the ATP concentration exceeds the physiological level of [ATP]=2000 $\mu$m, $\left\langle \Delta x_{\rm net}\right\rangle $ becomes almost insensitive to ATP concentration.
When the load $F$ is small, $\left\langle \Delta x_{\rm net}\right\rangle $ decreases sharply, demonstrating a notable sensitivity to small loads, as depicted in {\bf Fig.~S8(d)}. As the load $F$ increases, its effect on $\left\langle \Delta x_{\rm net}\right\rangle $ diminishes.

 According to the relationship between $V, \left\langle N\right\rangle , F/\left\langle N\right\rangle , L$ and $F$, we can naturally get {\bf Figs.~\ref{fit_F}(d-f)}.
Specifically, as the load $F$ increases, {\bf Figs.~\ref{fit_F}(d,e)} demonstrate that both the number of bound motors $\left\langle N\right\rangle$ and the force per motor $F/\left\langle N \right\rangle$ exhibit opposite trends to velocity. In contrast, {\bf Fig.~\ref{fit_F}(f)} demonstrates a consistent trend between the sliding distance and the velocity.

\begin{table}[b]
	\caption{\label{tab:1} Model parameter values obtained by fitting to experimental data of myosin II purified from muscles of frog measured in \cite{piazzesi2007skeletal}, see {\bf Fig.~\ref{fit_F}}. The $g_{\rm recovery}$ is limited by the presence of other parameters, resulting in the equation: $g_{\rm on}+g_{\rm ps}+g_{\rm off}+g_{\rm recovery}=-\Delta\mu_{\rm ATP}$.}
	\begin{ruledtabular}
		\begin{tabular}{lclc}
			\textrm{Parameter}&
			\textrm{Range tested}&	
			\textrm{Fitting value}\\
			\colrule
			$g_{\rm on}$ & $-10$ to 0 $k_{B}T$ \cite{wagoner2021evolution}& $-5.637$ $k_{B}T$ \\
			$g_{\rm ps}$& $-20$ to $-15$ $k_{B}T$ \cite{wagoner2021evolution}& $-17.438$ $k_{B}T$\\
			$g_{\rm off}$&$-10$ to 0 $k_{B}T$ \cite{wagoner2021evolution}&  $-3.499$ $k_{B}T$\\
			$g_{\rm recovery}$& $-$ & 1.574 $k_{B}T$\\
			$\omega _{\rm on}$&0.01 $-$ 10 s$^{-1}$ \cite{wagoner2021evolution}& 0.178 s$^{-1}$\\
			$\omega_{\rm off}$&0.1 $-$ 100000 s$^{-1}$ \cite{wagoner2021evolution}& 194.009 s$^{-1}$ \\
			$d^{\prime}$& 4 $-$ 8 nm \cite{kaya2017coordinated}& 4.187 nm \\
			$L_{p}$& 0.1 $-$ 1 nm \cite{fang2021biomechanical, li2005spectrin, fedosov2010systematic,oberhauser1998molecular}&0.162 nm\\
			$L_{c}$&9 $-$ 40 nm \cite{fang2021biomechanical, li2005spectrin, fedosov2010systematic,oberhauser1998molecular}& 24.561 nm \\	
			$k_{\rm pre}$& 0 $-$ 10000s$^{-1}$ \cite{kaya2017coordinated}& 5.03E$-06$ s$^{-1}$\\
			$k_{\rm mid}$& 0 $-$ 10000s$^{-1}$ \cite{kaya2017coordinated}&0.714 s$^{-1}$ \\
			$\gamma$& 0.1 $-$ 1 \cite{kaya2017coordinated,kaya2010nonlinear}& 0.192\\
		\end{tabular}
	\end{ruledtabular}
\end{table}
Since these biophysical quantities are related not only to the magnitude of external load $F$ but also to the ATP concentration, we provide the effect on each biophysical quantity as the ATP concentration varies, as shown in {\bf Fig.~S7(a)}.

The velocity increases with increasing ATP concentration because the motor cycle is accelerated, {\bf Fig.~S7(a)}. 
At very low ATP concentrations, the velocity is primarily determined by the slow detaching from the actin. When the concentration of ATP exceeds $10^{4}$, the rate of velocity ascent exhibits a tendency towards attenuation and velocity tends to flatten out,  although this is not shown in the {\bf Fig.~S7(a)}.

As ATP concentration increases, the average number of bound motors $\left\langle N\right\rangle $ decreases and becomes weakly dependent on load at high ATP concentrations. This means that the effect of load $F$ on $\left\langle N\right\rangle $ gradually decreases with increasing [ATP]. When the concentration of ATP decreases to 1 $\mu$m, the value of $\left\langle N\right\rangle$ tends to approach $N_{T}$. However, as ATP concentration approaches saturation, the downward tendency of $\left\langle N\right\rangle $ slows and essentially stabilizes, as shown in {\bf Fig.~S7(b)}. This is mainly because the increase in [ATP] directly leads to an increase in the rate $\omega_{\rm off}$, which further promotes the detachment of the motor from the actin filament, resulting in a decrease in $\left\langle N\right\rangle $.

The increase in force per motor $F/\left\langle N\right\rangle$ is a result of the decrease in $\left\langle N\right\rangle$ with increasing ATP concentration, while the external load $F$ is fixed. Moreover, the rate of increase in $F/\left\langle N\right\rangle$ is initially sluggish and then accelerates rapidly, as depicted in {\bf Fig.~S7(c)}.

The average net distance $\left\langle \Delta x_{\rm net}\right\rangle $ increases with ATP concentrations, but passes through a maximum and then decreases with further increasing [ATP], as shown in {\bf Fig.~S7(e)}. $\left\langle \Delta x_{\rm net}\right\rangle $ reaches its peak after the ATP concentration reaches 2000 $\mu$m. However, the degree of this increase in $\left\langle \Delta x_{\rm net}\right\rangle $ is reduced as $F$ increases, as shown in {\bf Fig.~S7(e)}. 
At low ATP concentrations, the effect of the external load $F$ on $\left\langle \Delta x_{\rm net}\right\rangle $ is not significant, and the values of $\left\langle \Delta x_{\rm net}\right\rangle $ are small. As the ATP concentration increases and reaches saturation, the influence of $F$ on $\left\langle \Delta x_{\rm net}\right\rangle $ becomes increasingly evident.
Specifically, as $F$ increases from 65 pN to 400 pN, the magnitude of $\left\langle \Delta x_{\rm net}\right\rangle $ increment becomes progressively less pronounced. At $F = 65$ pN, the greatest increase in $\left\langle \Delta x_{\rm net}\right\rangle $ is observed, reaching 0.49 nm. However, at $F = 400$ pN, the increase is minimal, with only 0.08 nm, as illustrated in {\bf Fig.~S7(e)}.

The trend of the sliding distance plateauing and subsequently decreasing with increasing ATP concentration, as shown in {\bf Fig.~S7(d)}. One possible reason for this is that the continuously increase in ATP concentration leads to a decrease in the number of bound $\left\langle N\right\rangle $.

\subsection{The energy efficiency and power output of myosin half-filament}
In this section, we predict the thermodynamic efficiency $\eta$ and power output $P$ of muscle contraction over different external load $F$ and [ATP].
The thermodynamic efficiency of the myosin half-filament is
\begin{equation}
	\eta=\frac{F\left\langle\Delta x_{\mathrm{net}}\right\rangle}{\Delta \mu_{\mathrm{ATP}}},
	\label{eta}
\end{equation}
where $F\left\langle \Delta x_{\text {net}}\right\rangle$ represents the average output work $W$ exerted on the actin filament during one complete {\it microstate} cycle \cite{wagoner2021evolution}. 

Furthermore, the power output is determined by the product of load $F$ and sliding velocity, given by $P=F\cdot V$.
\begin{figure}[htbp]
	\includegraphics[scale=0.324]{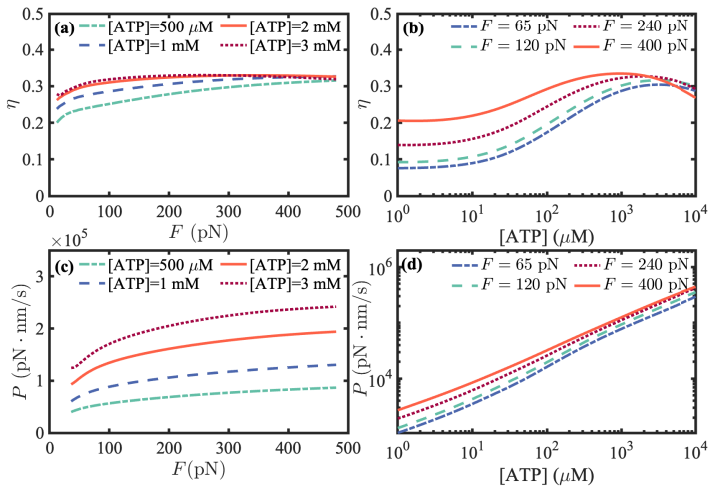}\\
	\caption{\label{effP}  The efficiency ($\eta$) and power output ($P$) of muscle contraction over different external load $F$ and [ATP].}
\end{figure}	
In {\bf Fig.~\ref{effP}(a)}, the efficiency increases as the external load $F$ increases and eventually approaches a limiting value of around 31\%. At low ATP concentrations, the effect of external load $F$ on efficiency is more pronounced, resulting in larger changes in efficiency, as shown in {\bf Figs.~\ref{effP}(a,b)}. After the ATP concentration reached saturation, the effect of $F$ change on efficiency is relatively less, and the rise in efficiency became more moderate. 
For instance, at an ATP concentration of 500 $\mu$M, increasing load $F$ caused efficiency to rise from 20.05\% to 31.57\%, corresponding to a change of 11.52\%. By contrast, at an ATP concentration of 2000 $\mu$M, the increase in efficiency with increasing $F$ is only 4.01\%, with efficiency rising from 27.88\% to 31.89\%, as shown in {\bf  Fig.~\ref{effP}(a)}.
 Furthermore, the efficiency curves are very close to each other and almost overlap at [ATP]=2000 $\mu$M and [ATP]=3000 $\mu$M.

In {\bf Fig.~\ref{effP}(b)}, the efficiency of myosin half-filament undergoes a transition from the load-sensitive regime (observed at low ATP concentrations) to the weak load-sensitive regime (occurring at higher ATP concentrations). Additionally, efficiency increases slowly at low ATP concentrations ranging from 1 to 10 $\mu$M. As ATP concentration increases from 10$\mu$M to 1000 $\mu$M, efficiency increases more rapidly until it reaches a maximum at ATP saturation, after which it begins to decrease. The position of the peak efficiency shifts towards lower ATP concentrations as $F$ increases. At $F=65$ pN, 120 pN, 240 pN, and 400 pN, the peak efficiencies are 30.48\%, 31.69\%, 32.71\%, and 33.41\%, respectively, occurring at ATP concentrations of 3162 $\mu$M, 3162 $\mu$M, 1995 $\mu$M, and 1000 $\mu$M, respectively.

The power exhibits a positive correlation with the increment in external load $F$, and its growth rate gradually slows down, as illustrated in {\bf Fig.~\ref{effP}(c)}. {\bf Fig.~\ref{effP}(d)} clearly depicts the exponential growth of power  with increasing ATP concentration. Additionally, power is affected by the velocity, which increases with rising ATP concentration when $F$ is fixed. The rate of power  increase tends to level off when the ATP concentration is above $10^{4}$.

\subsection {The steady-state distribution of myosin II motors}
	
\begin{figure}[htbp]
	\includegraphics[scale=0.315]{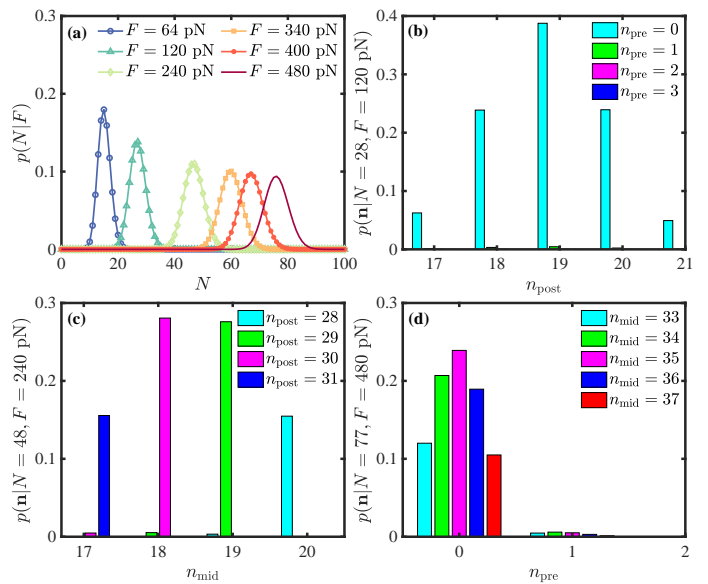}\\
	\caption{\label{probability}  The steady-state {\it mesostate} probability $p(N|F)$ and the {\it microstate}  conditional probability distribution $p({\bf n}|N, F)$ of myosin II motors. Since [ATP] is maintained constant at 2000 $\mu$m, we omit [ATP] here.  {\bf (a)} $p(N|F)$ versus the number of bound motors $N$ at different load $F$.  {\bf (b)} $p({\bf n}|N, F)$ of  $N$=28 at load $F$=120 pN versus the number of motors in {\it post-power stroke} state $n_{\rm post}$.  {\bf (c)} $p({\bf n}|N, F)$ of  $N$=48 at load $F$=240 pN versus the number of motors in {\it mid-power stroke} state $n_{\rm mid}$.  {\bf (d)} $p({\bf n}|N, F)$ of $N$=77 at load $F$=480 pN versus the number of motors in {\it pre-power stroke} state $n_{\rm pre}$.}
\end{figure}

Generally, when [ATP]=2000 $\mu$m and the external load $F$ is fixed, the steady-state {\it mesostate} probability $p(N|F)$ increases initially and then decreases with the number of bound motors $N$, forming a peak,  as shown in {\bf  Fig.~\ref{probability}(a)}. 
The distribution of $p(N|F)$ is mainly concentrated around the average number of bound motors $\left\langle N\right\rangle$. However, as the external load $F$ increases, the maximum value of this peak gradually decreases, and the shape of the distribution changes from tall and thin to short and wide.

When the external load $F$ is given, we identify the value of $N$ corresponding to the maximum steady-state {\it mesostate} probability $p(N|F)$, which corresponds to the peak of $p(N|F)$. For this particular $N$, there can be multiple sets of $n_{\rm pre}$, $n_{\rm mid}$, and $n_{\rm post}$ that satisfy the constraint $n_{\rm pre}+n_{\rm mid}+n_{\rm post}=N$. Therefore, this value of $N$ can correspond to multiple {\it microstate} configurations ${\bf n}$.
To further explore the distribution of {\it microstate} configurations ${\bf n}$ within this given $N$, we plot the conditional probability distribution $p({\bf n}|N,F)$ in two and three dimensions, as shown in {\bf Figs.~\ref{probability}(b-d)} and {\bf Figs.~S5(b-d)}.

When $F$=120 pN, the steady-state {\it mesostate} probability $p(N|F)$ reaches its maximum value at $N$=28. The corresponding conditional distribution $p({\bf n}|N$=28, $F$=120 pN$)$ is primarily concentrated in the range 17$\le$$n_{\rm post}$$\le$21, $ n_{\rm pre}$=0, and 7$\le $$n_{\rm mid}$$\le$11. 
This distribution exhibits an increasing and then decreasing trend in response to changes in the variable $n_{\rm post}$. The maximum probability value of $p({\bf n}|N$=28, $F$=120 pN$)$ is observed at the configuration $(n_{\rm pre}$=0, $n_{\rm mid}$=9, $n_{\rm post}$=19$)$, with a value of  0.39, as illustrated in {\bf Fig.~\ref{probability}(b)} and {\bf Fig.~S5(b)}.

Similarly, at external loads of $F$=240 pN and 480 pN, the steady-state probability $p(N|F)$ achieves its maximum values at $N$=48 and $N$=77, respectively. The conditional distribution $p({\bf n}|N$=48, $F$=240 pN$)$ primarily exhibits concentration in the range 28$\le$$n_{\rm post}$$\le$31, $n_{\rm pre}$=0, and $17$$\le$$n_{\rm mid}$$\le$20. Furthermore, $p(n|N$=77, $F$=480 pN$)$ is predominantly concentrated in the range 40$\le$$ n_{\rm post}$$\le$44, $n_{\rm pre}$=0, and 33$\le$$n_{\rm mid}$$\le$37.
The conditional distribution $p({\bf n}|N, F)$ increases and then decreases with changes in both $n_{\rm mid}$ and $n_{\rm pre}$. The maximum value of $p({\bf n}|N$=48, $F$=240 pN$)$ is 0.28 and occurs at $(n_{\rm pre}$=0, $n_{\rm mid}$=18, $ n_{\rm post}$=30$)$. 
For $p({\bf n}|N$=77, $F$=480 pN$)$, the maximum value is 0.24 and it occurs at $(n_{\text{pre}}$=0, $n_{\text{mid}}$=35, $n_{\text{post}}$=42$)$.
These results are shown in {\bf Figs.~\ref{probability}(c-d)}.

From {\bf Figs.~\ref{probability}(b-d)}, it is apparent that a large majority of the motors are in the {\it mid-power stroke} and {\it post-power stroke} states when $N$ motors are bound. Since motors in the {\it mid-power stroke} and {\it post-power stroke} states generate a relatively large force, having a larger number of motors in the {\it mid-power stroke} and {\it post-power stroke} states allows for more efficient sharing of the load.

In addition, with the increase in external load $F$, there is a shift in the predominant location $(n_{\rm pre}, n_{\rm mid}, n_{\rm post})$ where the conditional distribution $p({\bf n}|N, F)$ is concentrated, accompanied by an increase in the number of bound motors $N$. Meanwhile, the peak of the conditional distribution $p({\bf n}|N, F)$ decreases, as depicted in {\bf Figs.~\ref{probability}(b-d)} and {\bf Fig.~S5(b-d)}.

\subsection{Influence of power stroke on muscle performance}
\begin{figure}[htbp]
	\includegraphics[scale=0.375]{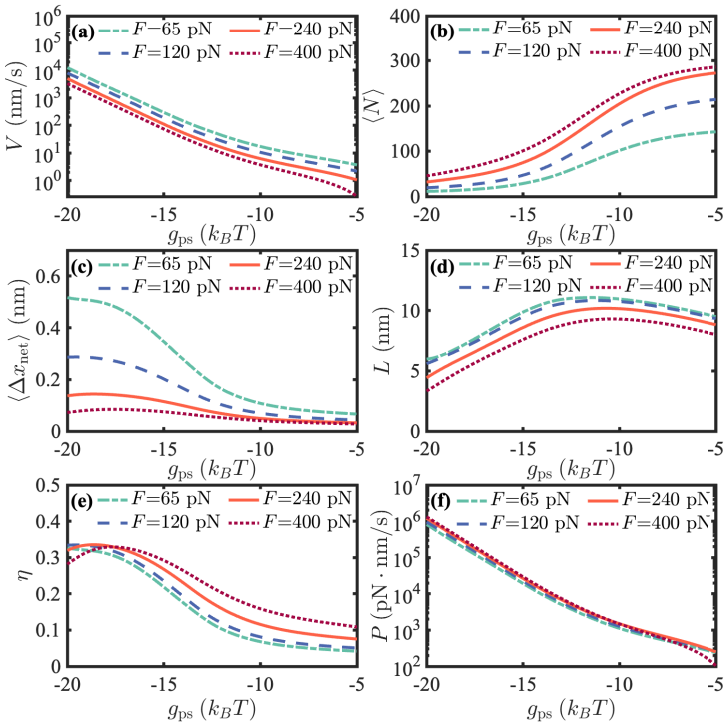}\\
	\caption{\label{GPS} The effects of variations in free energy bias $g_{\rm ps}$ on muscle performance at different values of external load $F$.
		\textbf{(a)} Velocity $V$.
		\textbf{(b)} Average number of bound motors $\left\langle N\right\rangle$. 
		\textbf{(c)} Average net distance $\left\langle \Delta x_{\rm net}\right\rangle $.
		\textbf{(d)} Sliding distance $L$.
		\textbf{(e)} Efficiency $\eta$.
		\textbf{(f)} Output power $P$. $\Delta g_{\rm ps} = g_{\rm ps}-g_{\rm ps0}$ and the $g_{\rm ps0}$ is fitting value  given in Tab.~\ref{tab:1}. In order to maintain the balance of basic free energy terms and ensure their sum remains equal to $-\Delta \mu_{\rm ATP}$, we evenly distribute the variation of $g_{\rm ps}$ by $\Delta g_{\rm ps}$ among the remaining basic free energy terms. $g_{\rm on}$, $g_{\rm off}$, and $g_{\rm recovery}$ undergo a variation of $-\Delta g_{\rm ps}/3$.}
\end{figure}	
Muscle contraction is initiated by the power stroke of myosin II motors, which generates force and motion. In addition, the strength of the power stroke is determined by the free energy bias between the {\it pre-power stroke} and {\it post-power stroke} states, $g_{\rm ps}$ \cite{wagoner2021evolution}. In simpler terms, ATP hydrolysis supplies the energy required for the power-stroke process, and the magnitude of this energy is represented as $|g_{\rm ps}|$.
Consequently, a more negative value of $g_{\rm ps}$ corresponds to a stronger power stroke, resulting in a greater proportion of motors being in the {\it mid-power stroke} and {\it post-power stroke} state.

To investigate the impact of $g_{\rm ps}$ on muscle performance, we set a range of values for $g_{\rm ps}$ from $-20$ $k_{B}T$ to $ -5$ $k_{B}T$, and we observe the trends in biophysical quantities with respect to different $g_{\rm ps}$ values, as shown in {\bf Fig.~\ref{GPS}}.

A more negative value of $g_{\rm ps}$ leads to an increase in the rates of $\omega_{\rm pre}$ and $\omega_{\rm mid}$, which accelerates the motor's cycle and ultimately results in an increase in velocity, as shown in {\bf Fig.~\ref{GPS}(a)}.
When [ATP] is 2000 $\mu$M, the total energy released by ATP hydrolysis remains constant, given by $g_{\rm on} + g_{\rm ps} + g_{\rm off} + g_{\rm recovery} = -25$. 
Consequently, a decrease in $|g_{\rm ps}|$ results in an increase in $|g_{\rm on}|$ and $|g_{\rm off}|$, thereby providing more energy for the motor to attach to and detach from the actin filament. This change influences the increase in $\omega_{\rm on}$ and the decrease in $\omega_{\rm on}^{\prime}$, but the rate $\omega_{\rm on}^{\prime}$ is significantly smaller than $\omega_{\rm on}$. Therefore, the ultimate effect is an increase in the attachment rate of the motor. This leads to an increase in the average number of bound motors $\left\langle N\right\rangle$, as shown in {\bf Fig.~\ref{GPS}(b)}.

{\bf Fig.~\ref{GPS}(c)} illustrates that as the absolute value of $g_{\rm ps}$ decreases, the average net distance $\left\langle \Delta x_{\rm net}\right\rangle $ of the motor also decreases. One possible explanation for this reduction is due to the increase in attached motors. The more attached motors, the shorter the average net distance  $\left\langle \Delta x_{\rm net}\right\rangle $, which is consistent with the conclusion of \cite{wagoner2021evolution}.

The peculiar trend observed in the sliding distance, as depicted in {\bf Fig.~\ref{GPS}(d)}, arises from the combined influence of the number of bound motor $N$ and the  net distance  $\Delta x_{\rm net}$.

In {\bf Fig.~\ref{GPS}(e)}, the efficiency exhibits an increasing trend with the absolute value of $g_{\rm ps}$ and gradually reaches a maximum as $g_{\rm ps}$ approaches the range of $-17$ $k_{B}T$ to $-19$ $k_{B}T$. However, beyond this range, the efficiency starts to decline as $|g_{\rm ps}|$ further increases. Importantly, our fitted value of $g_{\rm ps}$ at $-17.438$ $k_{B}T$ precisely falls within this range.

The power output exhibits an increase as the absolute value of $g_{\rm ps}$ rises, as shown in {\bf Fig.~\ref{GPS}(f)}. This trend of power can be attributed to the influence of velocity when force $F$ remains constant.

\subsection{The mean run time/length, mean existence probability and half-life period of myosin filament}

In this section, we investigate the mean run time/length, mean existence probability and half-life period of myosin half-filament, shedding light on actomyosin overall behavior.

\begin{figure}[htbp]
	\includegraphics[scale=0.34]{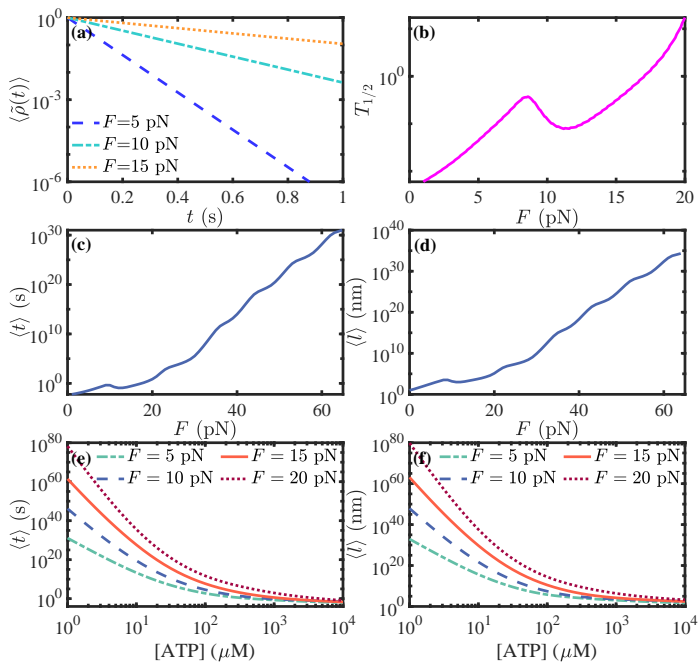}\\
	\caption{\textbf{(a)} Mean existence probability $\left\langle \tilde{\rho}(t) \right\rangle$. \textbf{(b)} Half-life $T_{1/2}$.  \textbf{(c)} Mean run time $\left\langle t\right\rangle $ as load $F$ varies at [ATP] = 2000 $\mu$M. \textbf{(d)} Mean run length $\left\langle l\right\rangle $ as load $F$ varies at [ATP] = 2000 $\mu$M. \textbf{(e)} Mean run time $\left\langle t\right\rangle $ as [ATP] changes at different loads. \textbf{(f)} Mean run length $\left\langle l\right\rangle $ as [ATP] changes at different loads.}
	\label{runtime6}
\end{figure}

{\bf Figs.~\ref{runtime6}(c,d)} demonstrate a rapid increase in the mean run time $\left\langle t \right\rangle $ and mean run length $\left\langle l\right\rangle $ of myosin filament as the load $F$ varies, with [ATP]=2000 $\mu$m. This implies that the motor remains almost constantly attached to actin, even at low loads (20 pN$\le$$F$$\le$ 60 pN), where the mean run time $\left\langle t\right\rangle$ is already considerably long.

{\bf Figs.~\ref{runtime6}(e,f)} illustrate that the mean run time $\left\langle t \right\rangle $ and mean run length $\left\langle l\right\rangle $ of myosin filament detachment from actin decrease monotonically with increasing [ATP]. At low ATP concentration, myosin II motors will stay on actin for more time, since the period of single cycle becomes long due to the lack of ATP molecule, namely $\omega_{\rm off}$ is small.  Furthermore, at low ATP concentrations, both the $\left\langle t \right\rangle $ and $\left\langle l\right\rangle $ are load-dependent, whereas they gradually become less dependent on force as the [ATP] approaches saturation.

In addition, we can evaluate the mean existence probability of myosin filament detaching from the actin filament after time $t$, denoted as $\left\langle \tilde{\rho}(t) \right\rangle$, as shown in Eq.~\eqref{meanrho}.
We can observe from the {\bf Fig.~\ref{runtime6}(a)} that the decay of the mean existence probability approximates an exponential decrease. 
However, the $\ln_{}{ \left\langle \tilde{\rho}(t) \right\rangle} $ curve is not a strictly linear one, because the second derivative of $\ln{\left\langle \tilde{\rho}(t) \right\rangle}$ decays rapidly and approaches zero.
Half-life $T_{1/2}$ refers to the time required for the mean existence probability $ \left\langle \tilde{\rho}(t) \right\rangle$ to decay to $1/2$, as shown in {\bf Fig.~\ref{runtime6}(b)}.

The curves in {\bf  Figs.~\ref{runtime6}(b-d)}  are not smooth, but rather exhibit fluctuations. One possible explanation for this is the discontinuity of $N_{T}$ caused by our use of integer values. These fluctuations are inherent in the $N_{T}$ model and cannot be avoided. Nevertheless, the overall trend of the curves is correct, indicating that the $\left\langle t \right\rangle $, $\left\langle l\right\rangle $ and $T_{1/2}$ are generally positively correlated with the load $F$. To ensure the accuracy of our findings, we have acknowledged the issue and conducted multiple parameter sets to validate our results.

\section{Discussion}
By integrating the mechanochemical model of a single myosin II motor with a cooperative model encompassing multiple myosin II motors, we comprehensively investigate both the {\it microstate} and {\it mesostate} dynamics of this myosin system. Our analysis reveals intriguing insights into the distribution of $p(N|F)$ in the steady state, which primarily concentrates around the average number of bound motors $\left\langle N\right\rangle $. As the load $F$ increases, the maximum value of the peak gradually diminishes, inducing a notable shift in the distribution shape from being tall and thin to becoming short and wide.

Significantly, a substantial fraction of motors bound to actin is found to reside in the {\it mid-power stroke} and {\it post-power stroke} states. This observation bears crucial implications as motors in the {\it mid-power stroke} and {\it post-power stroke} state possess the capability to generate relatively larger forces. Moreover, our nonlinear model uncovers their enhanced load-sharing capabilities. Consequently, this population of motors facilitates more efficient load sharing by augmenting the number of motors engaged in this particular state.

Additionally, we conducted a systematic analysis of the impact of load $F$ and [ATP] on the biophysical quantities of myosin II motors. By systematically exploring these two key factors, we aimed to gain deeper insights into the underlying mechanisms governing the behavior of myosin II in a complex and dynamic environment. 
 This contributes to a more comprehensive understanding of the regulatory mechanisms governing the functionality of this vital molecular motor system. 

Increasing the load $F$ results in a monotonic decrease in velocity. Meanwhile, the average number of motors $\left\langle N\right\rangle $ bound to actin increases, and $\left\langle N\right\rangle $ exhibits an increasing sensitivity to variations in [ATP]. Additionally, the sliding distance gradually decreases with increasing load $F$.

At near-vanishing load, the decrease in [ATP] leads to a rapid reduction in velocity, accompanied by a sharp decrease in the $\left\langle \Delta x_{\rm net}\right\rangle $. However, as the load $F$ increases, the rate at which velocity decreases gradually slows down, and the impact of load $F$ on $\left\langle \Delta x_{\rm net}\right\rangle $ diminishes.

With increasing [ATP], the velocity of the actin increases due to the acceleration of the motor cycle. At the same time, the $\left\langle N\right\rangle $ rapidly decreases and becomes weakly dependent on load $F$ at high ATP concentrations. 

$\left\langle \Delta x_{\rm net}\right\rangle $ initially increases with low ATP concentrations but reaches a maximum and then decreases with further increases in ATP concentration.  At low ATP concentrations, the effect of the load $F$ on $\left\langle \Delta x_{\rm net}\right\rangle $ is not significant. However, as ATP concentration increases and reaches saturation, the influence of load $F$ on $\left\langle \Delta x_{\rm net}\right\rangle $ becomes increasingly evident. Nevertheless, the degree of this $\left\langle \Delta x_{\rm net}\right\rangle $ increase diminishes as the load $F$ intensifies.
When the ATP concentration exceeds the physiological level of [ATP] = 2000 $\mu$m, $\left\langle \Delta x_{\rm net}\right\rangle $ becomes almost insensitive to ATP concentration while remaining highly dependent on the external load $F$.

Furthermore, we conducted an analysis to investigate the influence of $F$ and [ATP] on the efficiency and output power of myosin filament. The efficiency  shows a dependence on the load $F$, with an increase in load $F$ leading to an increase in efficiency. This trend continues until it reaches a limiting value of approximately 31\%, which is consistent with the model described by Wagoner $et.\ al.$ \cite{wagoner2021evolution}. 

In addition, the efficiency undergoes a transition from a load-sensitive regime at low ATP concentrations to a weak load-sensitive regime at higher ATP concentrations. 
Finally, our findings reveal that increasing the load $F$ or elevating the ATP concentration to saturation levels results in a substantial enhancement in both efficiency and output power.

Interestingly, as the $|g_{\rm ps}|$ is enhanced, it has two significant effects on the myosin system. Firstly, it results in an increase in velocity, reflecting the more forceful contraction of half-sarcomere. Secondly, it leads to a decrease in the average number of bound motor $\left\langle N\right\rangle $, suggesting that stronger power strokes make it more difficult for motors to attach to actin.

We observe an intriguing relationship between efficiency and the $g_{\rm ps}$. The efficiency shows an increasing trend with the $|g_{\rm ps}|$, indicating that a stronger power stroke generally improves the efficiency of the myosin filament. This trend continues until $g_{\rm ps}$ reaches the range of approximately $-17$ $k_{B}T$ to $-19$ $k_{B}T$, where the efficiency reaches a maximum value. However, beyond this range, further increases in $|g_{\rm ps}|$ lead to a decline in efficiency.
This finding highlights that enhancing the power stroke strength initially boosts efficiency, but there is a threshold beyond which the excessive power stroke strength becomes counterproductive and reduces efficiency.

Finally, the results of our study suggest that under physiological concentrations, the mean run time $\left\langle t\right\rangle $ and mean run length $\left\langle l\right\rangle $ of myosin filament exhibit a rapid increase as the load $F$ varies. This implies that as the load $F$ increases, the myosin II motors spend significantly longer periods attached to the actin filament.

On the other hand, the $\left\langle t\right\rangle $ and $\left\langle l\right\rangle $ show a monotonous decrease with increasing ATP concentration. Specifically, at low ATP concentrations, myosin filament tends to remain attached to the actin filament for longer durations, and both the $\left\langle t\right\rangle $ and $\left\langle l\right\rangle $ are load-dependent. However, as the ATP concentration approaches saturation, the dependence on load gradually decreases.
These observations may have significant implications for shedding light on the overall behavior of the myosin system and enhancing our understanding of it.

\section{Method}	
When the system enters the steady state, the biophysical quantities of myosin II can be obtained theoretically according to our  model, including velocity $V$, the average number of bound motors $\left\langle N\right\rangle $, and sliding distance $L$. 
	
The number of bound motors $N$ can be regarded as a linear Markov chain with {\it mesostate} rates, and the steady-state probability $p(N|F, {\rm [ATP]})$ can be obtained by using the master equation, we describe in {\bf Supplemental Material}, section F.
We can calculate the average number of bound motors as
	\begin{equation}
		\langle N\rangle=\sum_{N=1}^{N_{T}} Np(N|F, {\rm [ATP]}).
		\label{N}
	\end{equation}
For the same $N$, there are many cases of $n_{\rm pre}$, $n_{\rm mid}$ and $n_{\rm post}$, so one $N$ corresponds to multiple ${\bf n}$. The local equilibrium approximation solves $p({\bf n}|N, F, {\rm [ATP]})$ for the conditional distribution of all {\it microstates} within a given {\it mesostate} $N$ \cite{wagoner2021evolution}. The explicit expression for $p({\bf n}|N, F, {\rm [ATP]})$ is given in {\bf Supplemental Material}, section E.
Then, we can obtain $p \left( {\bf n}|F, {\rm [ATP]}) = p({\bf n}|N, F, {\rm [ATP]})p(N|F, {\rm [ATP]}\right) $.
	
The velocity-load relationship represents a steady state in which the motors repeatedly attach to, stroke and then detach from the actin.
In our model, the expression of velocity that sums the transition rates multiplied by the distance moved by the actin filament across all possible transitions of the cycle. Thus, the velocity of the actin filament is
	\begin{equation}
		\begin{aligned}
			V &= \sum_{\mathbf{n}} \Big[  \big( p(\mathbf{n}^{\prime\prime\prime})k_{\mathbf{n}^{\prime\prime\prime\prime},\mathbf{n}^{\prime\prime\prime}} - p(\mathbf{n}^{\prime\prime\prime\prime})k_{\mathbf{n}^{\prime\prime\prime}, \mathbf{n}^{\prime\prime\prime\prime}}\big)  \Delta x_{1}(\mathbf{n}^{\prime \prime \prime}) \Big] \\
			&\quad + \sum_{\mathbf{n}}\Big[ \big(p(\mathbf{n}^{\prime\prime\prime\prime})k_{\mathbf{n}, \mathbf{n}^{\prime\prime\prime\prime}} - p(\mathbf{n})k_{\mathbf{n}^{\prime\prime\prime\prime}, \mathbf{n}}\big)\Delta x_{2}(\mathbf{n}^{\prime\prime \prime \prime}) \Big]\\
			&\quad -\sum_{\mathbf{n}} \Big[ p(\mathbf{n})\sum_{i=n_{\rm pre}+n_{\rm mid}+1}^{N} \omega_{\mathrm{off}}(i, \mathbf{n})\Delta x_{\text{slip}}(i, \mathbf{n}) p(i) \Big]\\
			&\quad +\sum_{\mathbf{n}} \Big[ p(\mathbf{n}^{\prime})\sum_{k=n_{\rm pre}+n_{\rm mid}+1}^{N} \omega_{\text{on}}^{\prime}(k, \mathbf{n}^{\prime}) \Delta x_{\text{slip}}(k, \mathbf{n})p^{\prime}(k)\Big],
			\label{V}
		\end{aligned}
	\end{equation}
where $p\left(\mathbf{n}^{\prime\prime\prime}\right)=p\left(\mathbf{n}^{\prime\prime\prime}|F, {\rm [ATP]}\right)$ and similar for the other probabilities. $p(i)$ is the probability of  motor $i$ releasing from {\it post-power storke} state and $p'(k)$ signifies the probability of a motor in the {\it detached} state spontaneously extending to a sufficient extent to bind at position $k$ in the {\it post-power stroke} state. The distances $\Delta x_{1}\left(\mathbf{n}^{\prime \prime \prime}\right)$, $\Delta x_{2}\left(\mathbf{n}^{\prime\prime \prime \prime}\right)$, $\Delta x_{\text {slip}}(i, \mathbf{n})$, and $\Delta x_{\text {slip}}(k, \mathbf{n})$ are dependent on the load $F$, while $ \omega_{\mathrm{off}}(i, \mathbf{n})$ and $ \omega_{\text {on}}^{\prime}\left(k, \mathbf{n}^{\prime}\right)$ are influenced by both the load $F$ and [ATP], which we abbreviate here. 
Next, we obtain the expression for average net distance of  the actin filament during one full {\it microstate} cycle. The average net distance is
		\begin{align}
			\left\langle\Delta x_{\rm net}\right\rangle=&\sum_{\mathbf{n}}\Big [ \frac {p\left(\mathbf{n}^{\prime\prime\prime}\right)}{\sum_{\mathbf{n}}p\left(\mathbf{n}^{\prime\prime\prime}\right)}\Delta x_{1}\left(\mathbf{n}^{\prime \prime \prime}\right)\Big]\nonumber \\&+\sum_{\mathbf{n}}\Big [ \frac{ p\left(\mathbf{n}^{\prime\prime\prime\prime}\right)}{\sum_{\mathbf{n}}p\left(\mathbf{n}^{\prime\prime\prime\prime}\right)}\Delta x_{2}\left(\mathbf{n}^{\prime\prime \prime \prime}\right)\Big]	\label{Xnet1}\\
			&-\sum_{\mathbf{n}}\Big [ \frac{p\left(\mathbf{n}\right)}{\sum_{\mathbf{n}}p\left(\mathbf{n}\right)}\sum_{i=n_{\rm pre}+n_{\rm mid}+1}^{N}\Delta x_{\text {slip}}(i, \mathbf{n}) p(i)\Big].\nonumber
		\end{align}
	
Denoting by $F_N(t)$ the probability density that myosin filament separates from the actin (i.e., reaches 0 motor binding) for the first time at time $t$, starting from the state where there are $N$ motors binding at time $t=0$ where $N\in [1,2,\cdots,N_{T}]$.
It can be shown that ${\bf F}(t)=[F_{1}(t),\cdots, F_{N}(t),\cdots, F_{N_{T}}(t)]^{T}$ satisfy the following backward master equations, 
	\begin{equation}
		\frac{d {\bf F}(t)}{d t}=A {\bf F}(t)+[r(1)F_0(t),0,\cdots,0]^{T},
		\label{mean_run_time}
	\end{equation}
where $F_{0}(t)=\delta(t)$ in the first equation means that if myosin filament detaches from actin, the first-passage process is accomplished immediately, $r(1)$ is the detachment rate of only one motor from actin, and matrix $A$ is given in {\bf Supplemental Material}, section H.

The run time of myosin filament initiated with $N$ motors bound is $T_{N}=\int_{0}^{+\infty} t F_{N}(t) d t$.  So the mean run time of myosin filament along actin is
	\begin{equation}
		\left\langle T\right\rangle =\sum_{N=1}^{N_{T}} P\left( N|F\right) T_{N}.
	\end{equation} 
	And the mean run length of the myosin filament along actin is 
$	\left\langle l\right\rangle =	\left\langle T\right\rangle V.$
 
The existence probability can be represented as $\tilde{\rho}(t)={\bf e}-{\bf \rho}(t)=\int_{t}^{\infty}{\bf F}(t)dt=[\tilde{\rho}_{1}(t), \cdots, \tilde{\rho}_{N_{T}}(t)]^{T}$, see {\bf Supplemental Material}, section H. We can obtain the expression for existence probability 
	$\tilde{\rho}(t)=e^{At}{\bf e}$, with ${\bf {\tilde \rho}}(0)={\bf e}$. The mean existence probability  is then obtained by
	\begin{equation}
		\label{meanrho}
		\left\langle \tilde{\rho}(t) \right\rangle =\sum_{N=1}^{N_{T}} P\left( N|F\right) \tilde{\rho}_{N}(t).
	\end{equation}

	
\end{document}